\documentclass[doublecol]{epl2} 

\title{Delayed dynamic triggering of earthquakes: Evidences from a statistical model of seismicity}
\shorttitle{Dynamic triggering of earthquakes} 

\author{E. A. Jagla\inst{1} }
\shortauthor{E. A.  Jagla}

\institute{                    
  \inst{1} Centro At\'omico Bariloche, Comisi\'on Nacional de Energ\'{\i}a At\'omica, 
(8400) Bariloche, Argentina\\
}

\abstract{
I study a recently proposed statistical model of earthquake dynamics that incorporates aging as a 
fundamental ingredient. The model is known to generate earthquake sequences that quantitatively reproduce the spatial 
and temporal clustering of events observed in actual seismic patterns. The aim of the present work is to investigate if this model 
can give support to the empirical evidence that earthquakes can be triggered by transient small perturbations, 
particularly by the passing of seismic waves originated in events occurring in far geographical locations. 
The effect of seismic waves is incorporated into the model by assuming that they produce instantaneous
small modifications in the dynamical 
state of the system at the time they are applied. This change in the dynamical state has two main effects. On one side, 
it induces earthquakes that occur right at the application of the perturbation. These are called immediate events.
On the other side, after the application of the 
perturbation there is a delayed effect: the seismic activity increases abruptly after the 
perturbation, then falls down below the level of background activity, and eventually recovers to the background value. 
The time scale of these variations  depends on the internal dynamics of the system, and is totally independent of the duration of the perturbation. The number of delayed events in 
excess of the background activity is typically observed to be around a factor of twenty larger than the number of immediate events.
The origin of the enhanced activity period following the perturbation is associated to the existence of aging relaxation, and it 
does not occur if relaxation is absent. These findings give support to the experimental evidence that earthquake can 
be remotely triggered by small transient perturbations as those produced by seismic waves.}

\begin{document}

\maketitle

\section{Introduction}

Some earthquakes are triggered by previous earthquakes. The best known example is the 
occurrence of aftershocks, which are triggered by a previous large event, called the main shock. 
Aftershocks originate in the stress redistribution that the main shock produces around the rupture region\cite{dietrich}. This is why 
aftershocks occur typically inside or near the rupture region of the main shock\cite{scholz}. 
Aftershocks are an example of `static triggering', since the stress redistribution that generates them is permanent (at least until stresses are modified by
future events).

On the other hand, `dynamic triggering' refers to the possibility that some earthquakes are triggered by transient perturbations of the stress field, 
most remarkably by the passing of seismic waves produces by other earthquake \cite{freed}. 
Observational evidence of dynamic triggering has been accumulated since the nineties. For instance, the 1992 Landers earthquake triggered seismic activity 
as far as Yellowstone National Park \cite{hill}, and the 2002 Denali earthquake triggered seismicity in southeastern California \cite{prejean}.
Following these early observations, many evidences of dynamic triggering at large spatial distances have been accumulated (see \cite{freed} for a detailed set of references), in such a way that by now dynamic triggering of earthquakes is widely accepted to occur. Recently, evidences of dynamic triggering have been found also at the laboratory scale \cite{marone}.
Dynamically triggered events may occur right at the passage of the seismic waves\cite{hill,prejean,brodsky}. This is a case of `instantaneous'  triggering. However, one of the most intriguing characteristics of dynamic triggering is that in most cases the triggered events occur much later (hours or days) than the perturbation. This kind of `delayed' dynamic triggering is revealed by a comparison of the average seismicity before and after the passage of the seismic waves, and is undoubtedly documented for instance following the 1992 Landers event, which was seen to activate seismicity in a large region of western North America \cite{hill}.



The qualitatively most accepted explanation of dynamical triggering involves two steps. The first one corresponds to assume that the passage of seismic waves produces some kind of damage, or modification in the state of the fault, that contributes to `advancing the clock' of its seismic cycle\cite{gomberg}. The kind of physical modification that seismic waves can produce may involve for instance the increase in
pore fluid pressure, which leads to an effective decrease in effective normal stresses, favoring a weakening of the fault \cite{hill}. Another possibility 
for the physical effect of the passing seismic wave is the acceleration of subcritical crack growth \cite{atkinson}: a sudden, although transient, transfer of stress to the rock will cause the size of the cracks to increase some amount, moving the state of the fault towards a situation of instability.  
The second step in the dynamical triggering process is the completeness of the seismic cycle of the fault, leading to failure at an earlier time that it would have been in the absence of the perturbation. 
On a theoretical perspective, the process of advancing the clock of the fault and its eventual failure have been described in terms of the phenomenological rate-and-state equations modeling fault friction \cite{gomberg,gomberg2}, and the results obtained support the previous schematic description.

Beyond this rather qualitative understanding of delayed dynamic triggering, it would be important to have a model in which the described scenario can be
verified in its details, without introducing special hypothesis related to dynamic triggering itself. Statistical models, for instance spring-block, or similar kind of models, have in the past provided a solid basis for the understanding of basic features of seismic phenomena \cite{bk,langer,ofc,bak}. For instance, it was found that the broad distribution of earthquakes sizes observed, and accounted for by the Gutenberg-Richter law \cite{scholz}, can appear as an emergent collective behavior in a system consisting of a large number of elements having all the same typical size. Recently \cite{jagla_jgr,jagla_pre}, it has been shown how the
inclusion of internal relaxation, or aging in these models, expands greatly their possibilities, giving rise to sequences of events 
that display typical spatial and temporal correlations compatible with actual sequences, and also showing that relaxation can be considered as a physical 
ingredient that generates friction properties comparable with those experimentally observed.

The aim of the present work is to study the possibility of dynamic triggering of earthquakes in this kind of statistical model with relaxation. As in the interpretation of field results, there is here also the problem to define on safe grounds what the actual effect of the perturbation is on the state of the system. This will necessarily imply some ad hoc hypothesis to be added to the definition of the model. I will assume some kind of random damage (explained in detail below) caused by the passage of the seismic wave, and take this damaged configuration to be the starting point of a simulation that is compared with a situation in which the perturbation is not included. The main results that I obtain are the following: The perturbation triggers some immediate events (occurring right at the time when the perturbation is applied). In addition, during the temporal evolution after the application of the perturbation, a variation of the average seismicity rate in the system is clearly detected as compared to the case without perturbation. Seismic activity
is at a maximum right after the perturbation, decreases in time becoming lower than the background reference value, and then recovers to the reference value. The typical time scale of this process is comparable to the time scale of the aftershock cascades of the largest events observed in the system without perturbation, and this time scale has nothing to do with the time during which perturbation is applied. The time-integrated temporal density of events does not differ significantly between the cases with and without perturbation, and in this sense, the perturbation can be though of as advancing in time the occurrence of some events in the system, instead of generating new ones. By counting the events in the peak of over abundance with respect to background after the perturbation, this number is seen to be around a factor twenty larger than the number of immediate events triggered by  the perturbation, pointing to the fact that delayed events will be much more significant than immediate events during dynamic triggering. The perturbation does not alter the distribution of events in magnitude from the case without perturbation. A small perturbation produces a rather large effect in the system in the form of delayed events: a perturbation that affects 1\% of the sites of the system produces eventually delayed events involving about 20\% of all sites. 

In addition, in the range of parameters studied, the variations of the seismicity observed around the background value are
proportional to the intensity of the perturbation, namely the effect of the perturbation can be reasonably described as the superposition of the effects produced by the damage produced on individual sites by the perturbation.
Before presenting the results of the simulations, for completeness in the next section the basic assumptions of the model are explained.

\section{Brief recalling of the model, and the damage assumption for dynamical triggering}

I give here the basic definitions of the model, as it was presented in \cite{jagla_pre}. An alternative justification from a more physical starting point with essentially the same results can be seen in \cite{jagla_jgr}. The model can be considered to be an elaboration on the Olami Feder and Christensen
(OFC) model to describe a single planar fault \cite{ofc}. In the OFC model a real variable $f_i$ is defined on every node of a two dimensional square lattice, represented here by the index $i$. The value of $f_i$
is interpreted as the local value of the friction force between a sliding block and a rigid underlying surface. Tectonic loading at some velocity $V$ is applied by increasing the values of all $f_i$ linearly in time, i.e. $df_i/dt=V$. 
Each time one of the $f_i$ reaches a threshold value of 1, a local instability occurs and an unload process occurs. The corresponding $f_i$ is set to zero, and the neighbor sites $j$ are loaded according to $f_j \to f_j +\alpha f_i$, where $\alpha$ is a parameter between 0 and 1/4. This can produce some neighbor sites to overpass the threshold, and so a cascade can occur, until all $f_i<1$. This cascade is an earthquake, or an event in the system. 

The modifications introduced in \cite{jagla_pre} to the OFC model to obtain realistic features of earthquakes are the following: 1- thresholds are not uniform but they are site dependent, noted by $f^{th}_i$, and chosen from a Gaussian distribution with unitary mean and deviation $\sigma$ (in this paper I will use $\sigma=0.3$). Each time a site reaches the threshold and is downloaded, $f_i$ is decreased by 1, an amount $\alpha$ is uploaded onto each neighbor and the local threshold is chosen anew. The magnitude of each event in the system is calculated as $M=2/3 \log_{10} S$, where the `seismic moment' $S$ is the total number of topples that composed the complete avalanche.
2- The time evolution of $f_i$ between events is not simply governed by tectonic loading, there is now a relaxational term that tends to make the values of $f_i$ spatially uniform. The equation that describes temporal evolution of $f_i$ is
\begin{equation}
\frac{df_i}{dt}=V+R(\nabla^2 \nabla^2 f)_i
\label{dfdt}
\end{equation}
where $\nabla^2$ is the discrete Laplacian on the square lattice. This is one of the possible relaxation terms discussed in \cite{jagla_pre}. A physical motivation of its form can be found in \cite{jagla_jgr}. Note that the alternative of using a single Laplacian instead of the double one in eq. (\ref{dfdt})
produces only minor differences, as explained in \cite{jagla_pre}.
The relaxation parameter $R$ controls the intensity of the relaxation process. In addition, periodic boundary conditions are used.  
The model just described generates temporal sequences of events that display realistic aftershocks, produces a GR law with a correct decaying exponent without tuning of parameters, and displays realistic features of the friction process, as for instance the phenomenon of velocity weakening. 

On top of this model, I want to study the effect of the perturbation produced by the passage of seismic waves. These seismic waves are supposed to be generated at some remote location that is not included in the modeling. In particular, they do not originate in the earthquakes that are observed to occur
in the model. An additional hypothesis is necessary here, since we do not really know what the effect of the passage of seismic waves (or any other perturbation) is. But is not totally unreasonable to assume that the effect, whatever the actual physical mechanism is, can be described as a (small) change in the dynamical state of the system. I consider two possible modifications generated by the perturbation. They will be called model A and model B. In model A, at the time of the perturbation, a fraction $\epsilon$ of the sites of the system are chosen at random, and they are unloaded by the normal rules in the model. This may be interpreted as if there is a uniform probability $\epsilon$ of failure caused by the passage of the seismic wave.
In the second realization (model B), the sites that fail are those for which $f^{th}_i-f_i$ is lower than some threshold value $\Delta f$. This may be interpreted as the seismic wave producing the oscillation of the values of $f_i$ with some amplitude $\Delta f$. In this case, the intensity of the perturbation can be characterized also by the fraction $\epsilon$ of sites that became destabilized because of the perturbation. 

\section{Results}

Before going into the detailed results, it is convenient to emphasize that the delayed dynamic triggering to be observed in the model is associated to the existence of relaxation (i.e., a non zero value of $R$ in eq. (\ref{dfdt})). In fact, in the case in which $R=0$, the perturbation causes no variation of the average activity, except for the existence of some immediately triggered events (in an amount similar to those observed with relaxation), in particular, delayed triggered events are not detected for $R=0$.

\begin{figure}
\onefigure{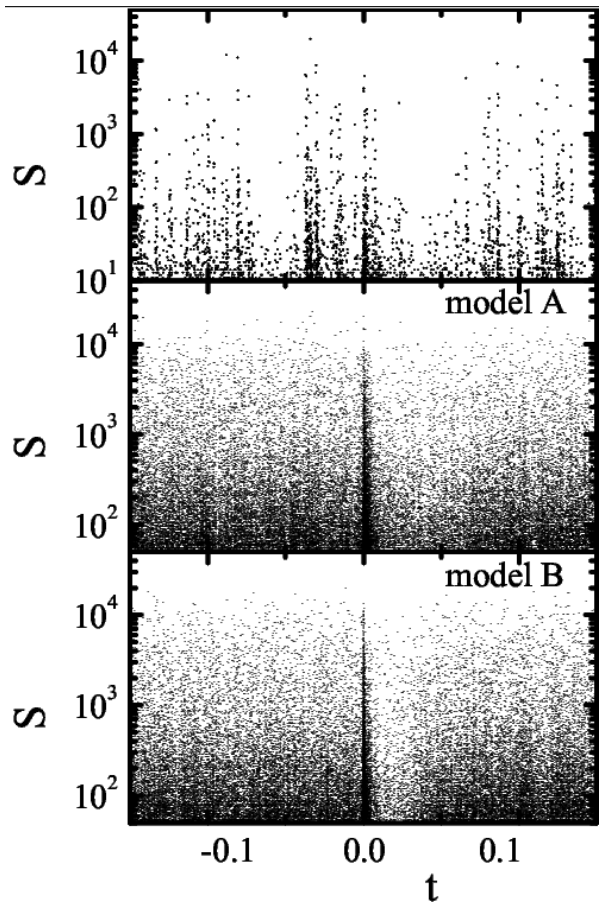}
\caption{(a) A single realization of the events in a system of size 256 $\times$ 256 ($R/V=50$, $\alpha=0.24$). At $t=0$ a perturbation (of type A) was applied, which destabilized a fraction $\epsilon=0.01$ of the sites in the system. This produced at the same $t=0$ cascades that involved a fraction of about 0.03 of all the sites in the system. These immediate events are not shown in the figure. (b) The result of accumulating 50 realizations like the one in panel (a). (c) The equivalent results accumulated during 50 realizations, now for a type B perturbation, with the same value of $\epsilon=0.01$. In (b) and (c) the overabundance of events following the perturbation is clearly discernible, as well as a broader region of depleted activity at longer times.
}
\label{f0}
\end{figure}

Dynamic triggering can be characterized as the difference in the seismic activity of the system when a perturbation is applied, compared to the case in which no perturbation was applied. In fig. \ref{f0}(a) we see an example of the temporal activity in  a system of 256 $\times$ 256 sites in which a perturbation was applied at $t=0$ with an intensity $\epsilon=0.01$ (i.e., one percent of the sites were instantaneously destabilized by the perturbation). It is clear from this figure that
due to the intrinsic fluctuations in seismic activity, we cannot extract any reliable conclusion on the influence of the perturbation on the dynamics.
A possibility to circumvent this problem would be to increase the system size, so the fluctuations in the seismic activity average out. But this would imply to work with prohibitively large (in terms of computational time) system sizes.
A more reasonable possibility is to consider many realizations in which perturbation is applied, and seismic response is presented as the average value over realizations. In fig. \ref{f0}(b) and (c) we see the result of accumulating many realizations for the two kind of perturbations discussed, namely model A and model B. Now a clear difference between activity with and without perturbation emerges, which is the phenomenon of dynamic triggering. To quantify in more detail distributions of events like those in fig. \ref{f0}(b-c), I will present results mainly for two quantities: temporal density of events $N(t)$, i.e., the number of events occurring per unit of time, and temporal density of stress drop $\sigma(t)$, i.e., the seismic moment of all events occurring per unit of time. For both quantities, a lower size cutoff $S_0$ will be indicated, i.e., only events with seismic moment $S>S_0$ will be considered. By definition, both $N(t)$ and $\sigma(t)$ are constant in time in the case of no perturbation. After a perturbation is applied, there will be changes in these quantities, which recover their asymptotic values at large times. 

\begin{figure}
\onefigure{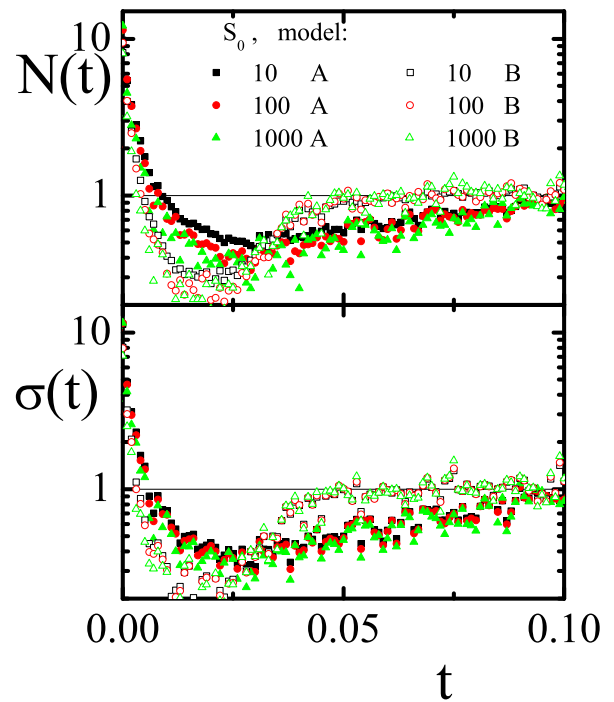}
\caption{Seismic activity as a function of time, averaged over about 400 realizations, following a perturbation occurring at $t=0$. Results for different values of the lower size cutoff $S_0$ are indicated, to show the independence on this cutoff. The results are normalized to the corresponding background activity. Immediate events are not displayed here, but they amount systematically to about 5\% of the number of delayed events (i.e., the peak above the background activity values)
}
\label{f2}
\end{figure}

In fig. \ref{f2} we see the results for $N(t)$ and $\sigma(t)$ for accumulated sequences like those in fig. \ref{f0}(b-c). The results are normalized to the background activity, and curves are presented for different values of the low size cut off $S_0$ used in counting events. 
The perturbation applied at $t=0$ brings about two main effects. First of all, there is a number of events that occur exactly at the time of the perturbation. They can be understood as the immediate cascades triggered by the sites that become unstable by the application of the perturbation. 
In addition, once the perturbation has disappeared and as time increases, the seismic activity is first enhanced, then it decreases, becoming even lower than the unperturbed value, to finally rejoin the unperturbed value at very long times. The events that form the peak above the unperturbed value will be referred to as the delayed dynamically triggered events. Their existence is in fact, the single most important result that I present in this paper. 
Within the precision of the curves in fig. \ref{f2}, the results are independent of the low size cutoff $S_0$ used to count the events, both for the immediate events and for the delayed ones. This is a general result that was observed in all simulations: dynamic triggering produces temporal variations of the seismic activity, but the distribution of events in magnitude continuous to be an intrinsic property of the model, independent of the perturbation. The time integrated values of both $N_t$ and $\sigma(t)$ are (again within the available precision) equal to the values without perturbation. For $\sigma(t)$ this is not a surprise, since on very large time intervals, the total stress drop can be calculated as the applied external stress on the system, and this is independent on the existence or not of a perturbation of the kind we are considering. For the case of $N(t)$ the fact was not obvious from the beginning. The fact that the number of events in the enhanced and depleted regions in fig. \ref{f2} are almost equal, comes out as a non-trivial results, which is actually related to the fact already mentioned that the distribution in magnitude of events is not altered. 

The fact that the perturbation does not change the total number of events allows a possible interpretation of the dynamical triggering in the following form: It can be said that dynamical triggering proceeds by making some events that were `scheduled' to occur later in time, to appear earlier,
because of the perturbation. While this interpretation is possible in principle, I think it is a bit dangerous to take it too seriously, as there is no clear comparison between the activity observed in a single realization of the dynamics with and without the perturbation. 

In comparing the results for model A and B in fig. \ref{f2}, we see that the results are qualitatively similar in both cases. A quantitative difference is that the time scale of the process is longer for model A, and shorter for model B. The reason of this behavior can be attributed to the fact that model A destabilizes a fraction of sites at random, whereas in model B the sites that are already close to instability are those effectively destabilized by the perturbation. This produces that for model $B$ a lower amount of tectonic loading ({\em i.e.} a shorter time) is necessary to erase the effect of the perturbation as compared to model A.

In addition, in a broad range of perturbation intensity studied, the effect of the perturbation is simply proportional to the perturbation intensity. In fact, results obtained for different values of the intensity $\epsilon$ of the perturbation (for $\epsilon$ between 0.001 and 0.02) show that the amount by which the results differ from those in the absence of perturbation scales linearly with the perturbation intensity. In particular, this implies that the time scale in which the delayed events are observed is independent of the perturbation intensity.
This linearity of the effect with the intensity of the perturbation, allows to make a concise statement on the number of events triggered by a perturbation.
In fact, the number of immediate events observed, amounts typically to roughly 5\%  of the number of dynamically triggered delayed events. This result is independent of the value of $S_0$ used as a lower size cutoff. Another way of stating this is to say that the eventual effect that a perturbation has onto the system is about a factor of 20 larger than what would be naively expected, i.e., than the effect observed immediately after the perturbation disappeared. This is an important result, because it points to the particular sensibility of the system to the perturbation, and indicates that (as it actually occurs in real seismicity) the largest amount of triggered events occur after a certain delay, and only a small fraction is instantaneous with the perturbation.

\begin{figure}
\onefigure{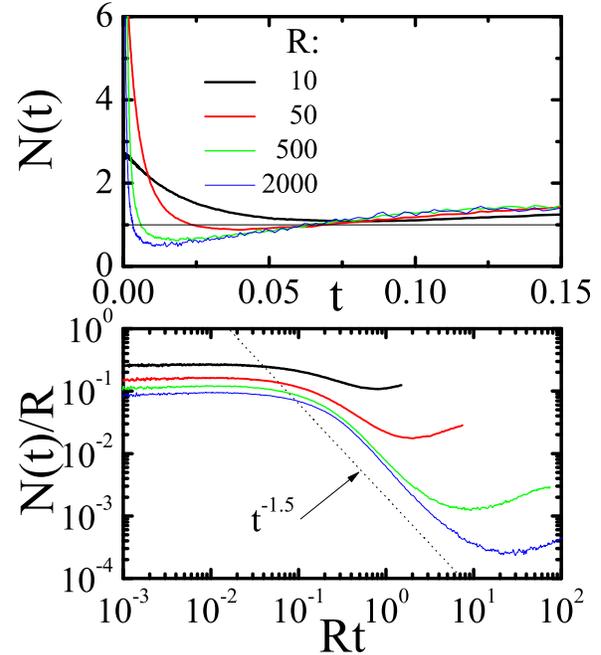}
\caption{Results for different values of the relaxation parameter $R$ in eq. (\ref{dfdt}) ($V=1$ and $\epsilon=0.01$ and $S_0=1$ in all cases, results for model A). In (b) the axis are scaled according to the value of $R$, and the data are presented in logarithmic scale. It can be seen that the peak of delayed dynamically triggered events scales with $R$ in such a way that its area is rather independent of $R$, particularly when $R$ becomes large, and the time span of these events is governed by $1/R$. Instead, the time required for the system to recover the background activity value is in all cases about 0.1, independently of $R$.
In (b), a temporal decay with an exponent $p=1.5$ is included, for comparison.}
\label{f3}
\end{figure}

Additional insight into the process of dynamical triggering is obtained by studying the response of systems with different intensity of relaxation $R$. First of all, it is clear from eq. (\ref{dfdt}) that a simultaneous change in $V$ and $R$ by the same factor is equivalent to a change in the time scale by the same factor, i.e., upon such a change in $V$ and $R$, results as those in fig. \ref{f2} would look exactly similar, with only the time axis rescaled. However, an independent change of $R$ and $V$ produces characteristic effects in the response of the system. For instance, in fig. \ref{f3} we see curves for different values of $R$, keeping $V$ as fixed. It can be seen that the temporal duration of the peak of dynamically triggered events is roughly proportional to $1/R$, but its area (i.e., the total number of dynamically triggered events) remains the same. On the other hand, the time that takes the system to recover the background activity is independent of $R$ (actually, it is proportional to $1/V$). 
The qualitative explanation of this behavior is the following: dynamically triggered events are produced by the relaxation mechanism (the $R$ term in eq. (\ref{dfdt})), and they would occur even in the case in which $V=0$. In this limit, it is clear that the effective time scale is inversely proportional to $R$, and this justifies the scaling of the peak of dynamically triggered events with $R$. On the other hand, once all dynamically triggered events have occurred, the system is in a state in which there has been an excess of seismic moment released. This deficit has to be eventually compensated by the loading mechanism, which is accounted for by the $V$ term. Thus we see that the temporal extent of regions of enhanced and depleted seismic activity are controlled by the independent parameters $R$ and $V$, respectively. In addition, if the ratio $R/V$ becomes sufficiently large,
the enhancement of seismic activity right after the perturbation will be much larger than the largest activity depletion, which becomes a very broad in time and quite shallow. This limit is compatible with actual seismic observations, as in this case there is no clear evidence of a period of depleted activity following the period of triggered activity.

There is some experimental evidence that the form of the time decay of excess activity caused by dynamic triggering follows approximately
the same Omori law that follow the usual aftershocks caused by static triggering. In the present model this is also the case. The decay of $N(t)$ may correspond to an Omori decay (fig. \ref{f3}(b)). However, for the ratios between $R$ and $V$ investigated, this decaying is masked at large times by the region of depleted activity, so at this point it can be only stated that the time decay observed of the peak of dynamically triggered events is not incompatible with the Omori law. A more definite statement could be obtained by analyzing cases with larger values of the ratio $R/V$.

\section{Conclusions}

In the present paper I have shown that the same mechanism that was used in Refs. \cite{jagla_jgr,jagla_pre} to obtain realistic sequences of earthquakes, is able to provide a consistent description for the phenomenon of dynamic earthquake triggering. A modification of the dynamical state of the model at some time $t=0$, assumed to be originated in the passage of seismic waves of some remote event, it was shown to produce in the model the following set of observations: A number of ``immediate" events at the time of the perturbation are triggered. An additional number of events (in excess of the normal activity) accumulate in the time after the perturbation has already vanished. These delayed dynamically triggered events are typically a factor of 20 larger than the number of immediate events. The time scale of delayed events is controlled by the relaxation parameter $R$ in the model, and in particular is independent of the duration of the perturbation (here it was assumed that perturbation occur during a vanishingly small time interval). The perturbation has no effect on the magnitude distribution of events.  The effect is proportional to the intensity of the perturbation, in a broad range of this intensity.

These findings give support to the somewhat debated phenomenon of dynamic triggering. The central point that the model captures, is the way in which a transient and small perturbation is able to trigger events that occur much later in time, once the perturbation has disappeared.



\acknowledgments
This research was financially supported by Consejo Nacional de Investigaciones Cient\'{\i}ficas y T\'ecnicas (CONICET), Argentina. Partial support from
grants PIP/112-2009-0100051 (CONICET, Argentina) and PICT 32859/2005 (ANPCyT, Argentina) is also acknowledged.

\end{document}